\documentclass[aps,twocolumn,showpacs,superscriptaddress,nofootinbib]{revtex4}

\usepackage{graphicx}
\usepackage{latexsym}

\usepackage{color}

\begin{document}
\title{ Driven Anomalous Diffusion: An example from Polymer Stretching}

\author{Takuya Saito}
\email[Electric mail:]{saito@fukui.kyoto-u.ac.jp}
\affiliation{Fukui Institute for Fundamental Chemistry, Kyoto University, Kyoto 606-8103, Japan}

\author{Takahiro Sakaue}
\email[Electric mail:]{sakaue@phys.kyushu-u.ac.jp}
\affiliation{Department of Physics, Kyushu University 33, Fukuoka 812-8581, Japan}

\def\Vec#1{\mbox{\boldmath $#1$}}
\def\degC{\kern-.2em\r{}\kern-.3em C}

\def\SimIneA{\hspace{0.3em}\raisebox{0.4ex}{$<$}\hspace{-0.75em}\raisebox{-.7ex}{$\sim$}\hspace{0.3em}} 

\def\SimIneB{\hspace{0.3em}\raisebox{0.4ex}{$>$}\hspace{-0.75em}\raisebox{-.7ex}{$\sim$}\hspace{0.3em}}

\date{\today}

\begin{abstract}
The way tension propagates along a chain is a key to govern many of anomalous dynamics in macromolecular systems. After introducing the weak and the strong force regimes of the tension propagation, we focus on the latter, in which the dynamical fluctuations of a segment in a long polymer during its stretching process is investigated.
We show that the response, i.e., average drift, is anomalous, which is characterized by the nonlinear memory kernel, and its relation to the fluctuation is nontrivial.
These features are discussed on the basis of the generalized Langevin equation, in which the role of the temporal change in spring constant due to the stress hardening is pinpointed. We carried out the molecular dynamics simulation, which supports our theory.
\end{abstract}

\pacs{36.20.Ey,87.15.H-,83.50.-v}

\def\degC{\kern-.2em\r{}\kern-.3em C}

\newcommand{\gsim}{\hspace{0.3em}\raisebox{0.5ex}{$>$}\hspace{-0.75em}\raisebox{-.7ex}{$\sim$}\hspace{0.3em}} 
\newcommand{\lsim}{\hspace{0.3em}\raisebox{0.5ex}{$<$}\hspace{-0.75em}\raisebox{-.7ex}{$\sim$}\hspace{0.3em}} 

\def\Vec#1{\mbox{\boldmath $#1$}}

\maketitle

\section{Introduction}
The fractional Brownian motion (fBm) is a class of stochastic process to describe the anomalous diffusion~\cite{Mandelbrot}. 
Being a natural extension of normal Brownian motion, it can be expressed in terms of its incremental sequence
\begin{eqnarray}
x(t) = \int_0^t ds \ \eta(s)
\end{eqnarray}
with $x(0)=0$, and the fractional Gaussian noise $\eta(t)$ with zero mean and the correlation
\begin{eqnarray}
\langle \eta(t) \eta(s) \rangle \sim |t-s|^{\alpha-2},
\end{eqnarray}
hence, the mean square displacement (MSD) $\langle x^2(t) \rangle \sim t^{\alpha}$, where $\langle \cdots \rangle$ represents the statistical average.

Many examples of fBm can be found in systems in cells and various soft matter systems. 
A list of Examples includes the motion of small particle (colloids, lipid granules, etc.) in cells and polymer networks~\cite{PRL_Tolic_2004,PRL_Amblard_1996}, the lateral diffusion inside lipid membranes~\cite{PRL_Akimoto_2011,PRL_Jeon_2012}, the process of polymer translocation through a nano-pore~\cite{PRL_Zoia_2009, JStatMech_Panja_2010_01, PRE_Dubbeldam_2011,JCP_deHaan_2012}, DNA/RNA hairpin formation~\cite{PRE_Walter_2012} and the telomere dynamics in the nucleus~\cite{PRL_Bronstein_2009}, etc. Here the visco-elastic response of the system is often invoked as a physical mechanism at hand to generate sub-diffusion ($0<\alpha<1$), where the noise $\eta(s)$ and the memory kernel $\mu(t)$ are related through the fluctuation-dissipation theorem (FDT)
\begin{eqnarray}
 \mu(t-s) k_BT = \langle \eta(t) \eta(s) \rangle
 \label{FDT_GLE}
 \end{eqnarray}
with $k_BT$ being the thermal energy (see, for instance~\cite{Jeon_Metzler, Sokolov} and references therein). 
In this case, the equation of motion describing the process is called the fractional Langevin equation, which can be written in the integral form as
\begin{eqnarray}
\frac{dx(t)}{dt} = \int_{-\infty}^t ds \ \mu(t-s) f(s) + \eta(t),
\label{GLE_1}
\end{eqnarray}
that is the generalized Langevin equation (GLE) with a power-law memory kernel, where $f(t)$ is an external force acting on the system.

 In this paper, we drive such a sub-diffusive walker by an external force and consider its fluctuating dynamics. 
Such a situation would be related to the active process in the biopolymer, such as polymer translocation driven by voltage drop~\cite{PRE_Sakaue_2007, EPL_Linna_2009, JPC_Rowghanian_2011, EPJE_Saito_2011, JCP_Ikonen_2012}, rotational dynamics of entangled polymers~\cite{EPJE_Walter_2014}, and chromosome segregation during cell division~\cite{NuclAcidsRes_Kuwada_2013, BJ_Lampo_2015}.
The quantities of interest are the time evolutions of the average displacement (AD) $\left< x(t) \right> $ and the variance in displacement (VD) $\left< \Delta x^2(t) \right> \ (=\mathrm{MSD}-\left< x(t) \right>^2)$, where $\Delta x(t) \equiv x(t)-\left< x(t) \right>$. 
When the memory kernel is independent of the driving force, an answer would be rather straightforward; the linear response theory suggests the anomalous drift $\langle x(t) \rangle \sim t^{\alpha} f$ and the fluctuation $\Delta x(t)$ around the average drift again becomes the fBm whose VD is given by $\langle \Delta x^2(t) \rangle \sim t^{\alpha}$ with the same anomalous exponent $\alpha$ as the undriven fBm. Then, the ratio $\left< \Delta x^2(t) \right>/\left< x(t) \right>$ becomes constant in time, which provide a way to evaluate the driving force from the time series of the trajectory (see Eq.~(\ref{x_xx}) below). The natural question here is how such a simple and potentially useful result should be altered (or not) beyond the linear response domain.  Indeed, the nonlinearity in the viscoelastic response would be pertinent to most of soft materials and biopolymers, where anomalous diffusion is commonplace. The memory kernel $\mu_f(t)$ and the noise $\eta_f(t)$, then, become force dependent.

As a model system, we analyze the motion of a tagged segment in a long polymer chain. This is indeed one of the paradigms for the sub-diffusion, where the GLE with a power-law memory kernel can be derived from a microscopic polymer model~\cite{JStatMech_Panja_2010_01,PRE_Sakaue_2013}. In our model, the nonlinearity arises from the self-avoidance (SA) and hydrodynamic interactions (HIs) between different segments in the polymer. Our main purposes are (i) to determine the nonlinear memory kernel $\mu_f(t)$ and (ii) to understand the nature of the fluctuation of the driven tagged segment determined by $\eta_f(t)$.
Though our main interest is in the nonlinear (strong force) regime, we also include the discussion for the linear response (weak force) regime. This not only makes the argument comprehensive, but also counterpoints the qualitative difference between these two regimes.

In Sec. II, we construct scaling theory for the anomalous dynamics of the tagged segment based on the notion of {\it time-dependent friction}. This will be done first for the weak driving regime (Sec. II A), and then for the strong driving regime (Sec. II B), where the response becomes generally nonlinear.
In Sec. III, we present molecular dynamics (MD) simulation results in the strong driving regime, which are discussed in light of the scaling theory. The result indicates that the analogous relation to Eq.~(\ref{FDT_GLE})
\begin{eqnarray}
\mu_f(t-s) k_BT = \langle \eta_f(t) \eta_f(s) \rangle
\label{FDT_GLE_f}
\end{eqnarray}
does not hold, and the fluctuation $\Delta x(t)$ is not necessarily described by fBm.
To get better insights into the anomalous driven dynamics, we proceed, in Sec. IV, the analysis of modes in a polymer. This enables us to derive the memory kernel from a microscopic polymer model. 
Again, we first give a paradigmatic treatment in the weak force regime (Sec. IV A,\ B), on which we develop an approximate argument suitably modified to the strong force regime (Sec. IV C). 
In Sec. V, we conclude this study.

\begin{figure}[t]
\begin{center}
\includegraphics[scale=0.80]{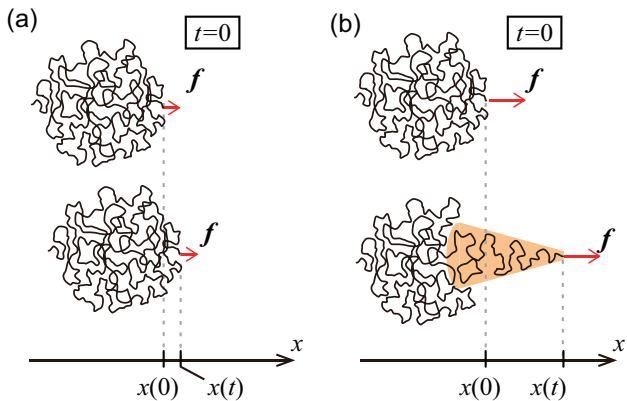}
      \caption{
	  (Color Online) Schematic representation of driven dynamics in polymer stretching in (a) near equilibrium and (b) strongly driven dynamics.
	  }
\label{fig1}
\end{center}
\end{figure}

\section{Scaling theory}
\label{scaling_theory}
The polymer consists of $N$ segments, each of which is characterized by its size $a$ and friction coefficient $\gamma$.
We prepare the polymer in its thermal equilibrium state, so that its average spatial size is
\begin{eqnarray}
R \simeq a N^{\nu}.
\label{R_eq}
\end{eqnarray}
The Flory exponent $\nu$ characterizes the equilibrium coil size; $\nu=1/2$ for the ideal chain; $\nu=3/4$ or $\nu \simeq 0.588$ in two-dimensional (2D) or 3D spaces, respectively, for the chain with SA~\cite{deGennesBook}. In addition, we write the longest relaxation time $\tau$ of the polymer as
\begin{eqnarray}
\tau \simeq \tau_0 \left(\frac{R}{a}\right)^{z},
\label{tau_eq}
\end{eqnarray}
where $\tau_0 \simeq \gamma a^2/k_BT$ is the segmental time scale, and the so-called dynamic exponent takes $z=3$ in the HIs dominated case (non-draining polymer) or $z=2 + (1/\nu)$ for the free-draining polymer~\cite{deGennesBook}.

\subsection{Time-dependent friction}
\label{time_dependent_friction}
Suppose that we switch on a constant external force $f$ acting on the end segment, hereafter called tagged segment, at $t=0$. We set its initial position as origin $x(0)=0$ (see Fig.~1(a)).  
As time goes on, the motion of the segment creates the tension, which gets transmitted along the polymer~\cite{PRE_Sakaue_2007,PRE_Sakaue_2012}. 
This leads to the {\it time-dependent friction} associated with the growing section of the correlated motion, thus gives rise to the memory effect to the motion.
Below, we construct the scaling form of the memory kernel from the force balance argument during the process. But the way the above scenario comes into effect depends on whether the motion of the segment is dominated by the thermal fluctuation or the driving force.
We are thus led to distinguish weak and strong force regimes.
For simplicity, we look at the position $x(t)$ of the segment in the direction along the force only; the motion in other directions is just an unforced fBm. 

{\it Weak force regime---.}
The force magnitude $f$ is weak enough $f < k_BT/R$ so that the polymer shape is kept in equilibrium coil.
In this weak force (near equilibrium) regime, the motion of the segment is essentially relaxational in thermal fluctuation. 
Let $r^*(t)$ be the distance, i.e., the root VD (or root MSD), that the segment travels during the time interval $t$. In the time window $t < \tau$, the polymer as a whole has no time to relax. This implies that the restoring force $\simeq k_BT/r^*(t)$ acts to the tagged segment, and at the same time, the tension due to the motion is transmitted up to $n^*(t) \simeq (r^*(t)/a)^{1/\nu}$ segments apart along the chain.
These $n^*(t)$ segments would take part in the motion of the tagged segment, thus, contribute to the friction $\gamma(t)$.
Therefore, the force balance equation reads
\begin{eqnarray}
\gamma(t) \frac{dr^*(t)}{dt} \simeq \frac{k_BT}{r^*(t)},
\label{force_balance_eq}
\end{eqnarray}
where the time-dependent friction is given by
\begin{eqnarray}
 \gamma(t) \simeq \gamma \left( \frac{r^*(t)}{a} \right)^{z-2}
\label{gamma_eq}
\end{eqnarray}
as inspected from Eq.~(\ref{tau_eq}).
Solving the differential equation~(\ref{force_balance_eq}), we obtain
\begin{eqnarray}
 r^{*}(t) (\simeq \sqrt{\langle \Delta x^2(t)\rangle}) \simeq  a \left( \frac{t}{\tau_0}\right)^{1/z}
 \label{r*_eq} \\
 n^{*}(t) \simeq \left( \frac{t}{\tau_0}\right)^{1/\nu z}, \label{n*_eq}
\end{eqnarray}
where $\Delta x(t) \equiv x(t)-\langle x(t) \rangle$.
We thus identify the anomalous diffusion exponent $\alpha = 2/z$, whose physical origin is the memory effect associated with the tension transmission. Superimposed on this diffusion is the (small) drift $\langle x(t) \rangle$, which can be obtained from the relation $\gamma(t) d\langle x(t)\rangle/dt = f$ as;
\begin{eqnarray}
\langle x(t) \rangle  \simeq a\left(\frac{fa}{k_BT}\right) \left( \frac{t}{\tau_0}\right)^{2/z},
\label{r_drift_eq}
\end{eqnarray}
where $\gamma(t)$ is determined from Eqs.~(\ref{gamma_eq}) and~(\ref{r*_eq}).
Comparing Eqs.~(\ref{r*_eq}) and~(\ref{r_drift_eq}), one finds that the drift is indeed negligible as long as $t \ll \tau_{f0}$ with
\begin{eqnarray}
\tau_{f0} \simeq \tau_0 \left( \frac{\xi}{a}\right)^{z} \simeq \tau_0 \left( \frac{fa}{k_BT}\right)^{-z}.
\label{tau_{f0}}
\end{eqnarray}
Given the condition $f< k_BT/R$, we find $\tau_{f0} > \tau$, ensuring the fluctuation dominance in the weak force regime.

{\it Strong force regime---.}
Now the force is strong enough ($f > k_BT/R$) to make the drift dominant over the diffusion for the tagged segment motion. At the same time, the polymer is stretched in the pulled direction. The resultant steady-state conformation can be pictured as a succession of blobs with growing size~\cite{EPL_Brochard_1993}. 
For simplicity of the argument, we here adopt the mono-block approximation~\cite{Macromolecules_Marciano_Brochard_1995}, i.e., the size $\xi$ of the blobs are uniform and given by $\xi \simeq k_BT/f$, which suffices for the scaling discussion (see Appendix A and e.g. Ref~\cite{PRE_Sakaue_2012} for the argument with spatial inhomogeneity).
The stretched length of the polymer is estimated as
\begin{eqnarray}
R_{\parallel} \simeq \xi \left( \frac{N}{g}\right) \simeq a \left( \frac{fa}{k_BT}\right)^{(1-\nu)/\nu}N,
\label{R_parallel}
\end{eqnarray}
where $g \simeq (\xi/a)^{1/\nu}$ is the number of segments inside the blob.

Again we switch on the force at $t=0$, before which the polymer assumes an equilibrium state at rest.
In the strong force regime, the driving force overwhelms the thermal fluctuation, thus governs the motion of the tagged segment. Therefore, we set $r^*(t)$ to be a drift distance traveled by the segment during the time interval $t$. The polymer as a whole has no time to react to the force, and the tension gets transmitted only up to $n^*(t) \simeq (r^*(t)/a)(fa/k_BT)^{(\nu-1)/\nu}$ segments apart from the tagged segment (e.g. Eq.~(\ref{R_parallel})). These segments follow the driving force, hence, the time-dependent friction grows as
\begin{eqnarray}
\gamma_f(t) \simeq \gamma \left( \frac{\xi}{a}\right)^{z-2}\frac{n^*(t)}{g} \simeq \gamma \left( \frac{r^*(t)}{a}\right)\left( \frac{fa}{k_BT}\right)^{3-z},
\label{gamma_f_t}
\end{eqnarray}
where we assume the equilibrium formula~(\ref{gamma_eq}) is valid up to the length scale $\xi$, which adds up in the larger scale. 
The force balance equation thus reads
\begin{eqnarray}
\gamma_f(t) \frac{dr^*(t)}{dt} \simeq f,
\label{force_balance_f}
\end{eqnarray}
the solution of which is
\begin{eqnarray}
 r^{*}(t) (\simeq \langle x(t)\rangle) \simeq  a \left( \frac{t}{\tau_0}\right)^{1/2}\left( \frac{fa}{k_BT}\right)^{(z-2)/2}
 \label{r*_f} \\
 n^{*}(t) \simeq \left( \frac{t}{\tau_0}\right)^{1/2}\left( \frac{fa}{k_BT}\right)^{(z/2)-(1/\nu)} \label{n*_f}. 
\end{eqnarray}
In addition to the anomalous drift exponent $1/2$, there arises the force dependence with the characteristic exponent $(z-2)/2$, i.e., the response is nonlinear except for the Rouse model $z=4$~\cite{PRE_Sakaue_2012,PRE_Rowghanian_Grosberg_2012}.

Superimposed on this drift is the (small) diffusion $\sqrt{\langle \Delta x^2(t) \rangle}$, the property of which is to be unveiled. One might estimate it from the analogous relation as Eq.~(\ref{force_balance_eq}); $\gamma_f(t) d \sqrt{\langle \Delta x^2(t) \rangle}/dt \simeq k_BT/\sqrt{\langle \Delta x^2(t) \rangle}$, which yields a conjecture
\begin{eqnarray}
\sqrt{\langle \Delta x^2(t) \rangle} \simeq a \left( \frac{t}{\tau_0}\right)^{1/4} \left( \frac{fa}{k_BT}\right)^{(z/4)-1}
\label{DX_f}
\end{eqnarray}
where $\gamma_f(t)$ is determined from Eqs.~(\ref{gamma_f_t}) and~(\ref{r*_f}).

{\it Remarks---.} 

(i) The effect of the force is a weak perturbation in the scale smaller than $\xi \simeq k_BT/f$. The corresponding time scale is $\tau_{f0} \simeq \tau_0 ( \xi/a)^{z}$ given in Eq.~(\ref{tau_{f0}}), which signals the onset time of the strong force regime. In the time range $t < \tau_{f0}$, the weak force regime applies~\cite{PRE_Sakaue_2007,PRE_Saito_2013}.

(ii) The terminal time of the driven process is given by the condition $r^*(\tau_f) \simeq R_{\parallel}$;
\begin{eqnarray}
\tau_f \simeq \tau_0 N^2 \left( \frac{fa}{k_BT}\right)^{(2/\nu)-z} \simeq \tau_{f0} \left( \frac{N}{g}\right)^2.
\label{tau_f}
\end{eqnarray}
At this time, the tension caused by the external force reaches to the other chain end.
This crossovers to the equilibrium formula~(\ref{tau_eq}) at $f \rightarrow k_BT/R$.

\subsection{Generalized Langevin Equation}
The above scaling arguments for the memory effect can be generalized to the case of the arbitrary protocol of the time-dependent driving force $f(t)$. This leads to the GLE given in Eq.~(\ref{GLE_1}) 
for the motion of the tagged segment~\cite{PRE_Sakaue_2013}.
The equivalent expression is
\begin{eqnarray}
0 = - \int_{-\infty}^t ds \ \Gamma(t-s) \frac{dx(s)}{ds} + f(t) + \omega(t),
\label{GLE_2}
\end{eqnarray}
where the friction kernel $\Gamma(t)$ and the noise $\omega(t)$ are related to the mobility kernel $\mu(t)$ and the noise $\eta(t)$ in Eq.~(\ref{GLE_1})  as ${\hat \Gamma} ({\hat z}) {\hat \mu} ({\hat z})=1$ and ${\hat \omega}({\hat z}) = {\hat \Gamma} ({\hat z}) {\hat \eta}({\hat z})$ in the Laplace domain~\cite{JStatMech_Panja_2010_01}.

{\it Weak force regime---.}
In our protocol, we switch on a constant force at $t=0$, i.e., $f(t)=f\Theta (t)$ with $\Theta (t)$ being the Heaviside step function. The time-dependent friction is then $\gamma(t) = [\int_0^t ds \ \mu(s)]^{-1}$. Using Eqs.~(\ref{gamma_eq}) and~(\ref{r*_eq}), we obtain
\begin{eqnarray}
\mu(t) &\simeq& - \left(\frac{1}{\gamma \tau_0} \right) \left( \frac{t}{\tau_0}\right)^{(2/z)-2}
\label{mu_eq} \\
\Gamma(t)&\simeq&  \left(\frac{\gamma}{ \tau_0} \right) \left( \frac{t}{\tau_0}\right)^{-(2/z)},
\label{Gamma_eq}
\end{eqnarray}
where the minus sign in $\mu(t)$ comes from the fact $2-z <0$ in practice. This reflects the viscoelastic response of the tagged segment leading to the sub-diffusion. Note that such a power-law regime should be viewed as an intermediate asymptotics valid in the time range $\tau_0 \ll t \ll \tau$ (See Sec.~\ref{ModeAnalysis}).
We assume that the memory kernels ($\mu$ and $\Gamma$) are related to the noises ($\eta$ and $\omega$) through FDT given by Eq.~(\ref{FDT_GLE}); the equivalent expression is $\Gamma(t-s)k_BT = \langle \omega(t) \omega(s) \rangle$.
The AD and VD can be calculated as
\begin{eqnarray}
\left< x(t) \right> &=& \int_0^t ds \int_0^s du \, \mu(s-u) f \simeq a\left(\frac{fa}{k_BT}\right) \left( \frac{t}{\tau_0}\right)^{2/z}\\
\left< \Delta x^2(t) \right>&=&\int_0^t ds \int_0^t du \, \langle \eta(s) \eta(u) \rangle \simeq  a^2 \left( \frac{t}{\tau_0}\right)^{2/z}.
\end{eqnarray} 
in agreement with Eqs.~(\ref{r_drift_eq}) and~(\ref{r*_eq}), respectively.

{\it Strong force regime---.}
Assuming the same line of argument as in the weak force regime, we obtain the estimate of the nonlinear memory kernel
\begin{eqnarray}
\mu_f(t) &\simeq& - \frac{1}{\gamma \tau_0}\left( \frac{t}{\tau_0}\right)^{-3/2}\left( \frac{fa}{k_BT}\right)^{(z/2)-2} \label{mu_f_scaling} \\
\Gamma_f(t) &\simeq& \frac{\gamma}{\tau_0}\left( \frac{t}{\tau_0}\right)^{-1/2}\left( \frac{fa}{k_BT}\right)^{2-(z/2)}.
\label{gamma_f_scaling} 
\end{eqnarray}
One can easily check that this yields the drift scaling, which is in accord with Eq.~(\ref{r*_f}).
The fluctuation around this drift is subtle, however. In Ref.~\cite{PRE_Sakaue_2013}, it was assumed that the property of the noise $\eta_f$ (or $\omega_f$) is encoded in the kernel through the relation~(\ref{FDT_GLE_f}).
With this naive assumption, the solution of the GLE leads to the VD scaling, which is in accord with Eq.~(\ref{DX_f}).
We repeat once more that this estimate needs to be checked.

{\it Summary---.}
The result and conjecture obtained so far based on the scaling argument are summarized as follows:
The leading component in the dynamics of the tagged segment is fluctuation or drift in weak or strong force regime, respectively. In either case, the motion of the tagged segment creates the tension, which gets transmitted along the chain with characteristic dynamics in respective regime. The resultant anomalous dynamics can be expressed using the memory kernel as
\begin{eqnarray}
\langle \Delta x^2(t)\rangle  &\simeq&  \frac{k_BT}{\Gamma (t)}  \qquad [{\rm weak \ force \ regime}]\\
\langle x(t) \rangle &\simeq& \frac{f}{\Gamma_f(t)} \label{X_f}\qquad [{\rm strong \ force \ regime}]
\end{eqnarray}
The FDT~(\ref{FDT_GLE}) or its analogous relation~(\ref{FDT_GLE_f}) then implies 
\begin{eqnarray}
\langle x(t) \rangle  &\simeq&  \frac{f}{\Gamma (t)}  \qquad [{\rm weak \ force \ regime}]\\
\langle \Delta x^2(t)\rangle &\simeq& \frac{k_BT}{\Gamma_f(t)} \qquad [{\rm strong \ force \ regime}] \label{DX2_f}
\end{eqnarray}
for the minor component in the motion.
In addition, it is worth pointing out that GLE formalism with FDT~(\ref{FDT_GLE}) or its analogue~(\ref{FDT_GLE_f}) predicts the quantitative relation between the drift and the fluctuation
\begin{eqnarray}
\left< \Delta x^2(t) \right> = \frac{2k_BT }{f}\left< x(t) \right>
\label{x_xx}
\end{eqnarray}
regardless of the specific form of memory kernel.
Very recently, this relation has been used to evaluate the driving force of the bacterial chromosome segregation {\it in vivo}~\cite{BJ_Lampo_2015}.

\section{Molecular dynamics simulations}
\label{MD}
\begin{figure}[t]
\begin{center}
\includegraphics[scale=0.70]{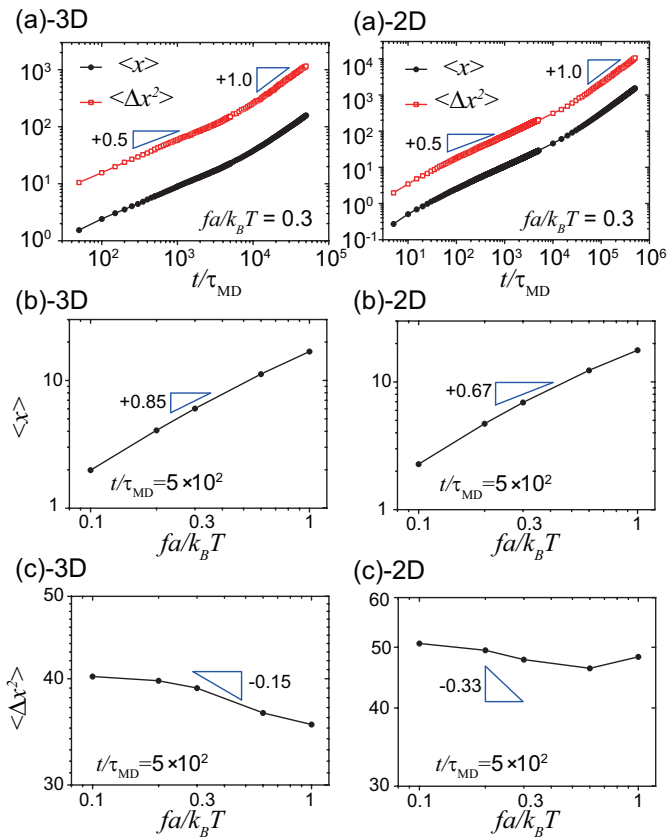}
      \caption{
	  (Color Online) 
	  Dynamics of the tagged segment from MD simulations.
	  Left or right column shows results in 3D or 2D, respectively.
	  (a) Time evolution of average drift and VD at $fa/k_BT = 0.3$. Force dependence of (b) average drift and (c) VD at time $t/\tau_{MD}=5\times 10^2$.  
	  All plots are displayed in a double-logarithmic scale.
	  Inset triangle slopes indicate the theoretical scaling exponents speculated by Eqs.~(\ref{gamma_f_scaling}),\,(\ref{X_f}),\,(\ref{DX2_f}).
	  }
\label{fig2}
\end{center}
\end{figure}

\begin{figure}[t]
\begin{center}
\includegraphics[scale=0.65]{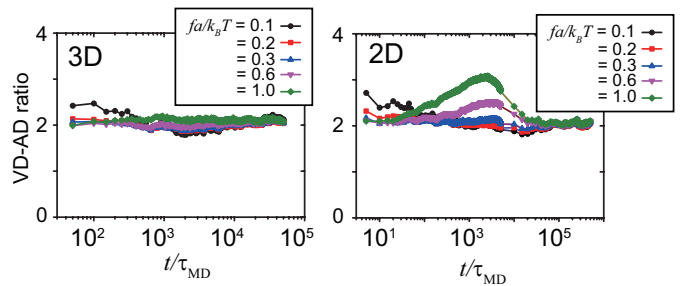}
      \caption{
	  (Color Online) 
	  VD-AD ratio $f \left<\Delta x^2(t) \right> / (k_BT\left< x(t) \right>) $ obtained from MD simulations in 3D (left) and 2D (right) as a function of time.
	  }
\label{fig3}
\end{center}
\end{figure}

In this section, we perform MD simulation to verify the scaling predictions in Sec.~II, in particular those Eqs.~(\ref{X_f}) and~(\ref{DX2_f}) in strong force regime.
In simulations, equation of motion for each segment is
\begin{eqnarray}
m \frac{d^2 \Vec{x}_i}{dt^2}=-\gamma \frac{d\Vec{x}_i}{dt} -\nabla_{\Vec{x}_i} U +\Vec{\zeta}_i(t) +f \delta_{iN} \Vec{e}_x,
\end{eqnarray} 
where $i (1 \sim N)$ is bead indices, $m$ and $\gamma$ are mass and frictional coefficient of a bead, $\Vec{\zeta}_i(t)$ is a Gaussian white noise with mean zero and the variance $\left<\Vec{\zeta}_i(t) \Vec{\zeta}_j(t') \right>=2\gamma k_BT \delta_{ij}\delta(t-t') \Vec{1}$ with $\Vec{1}$ being a unit matrix, and $\Vec{e}_x$ is the unit vector directed to $x$-axis.
HIs are ignored for simplicity (free draining).
The total potential $U=U_\mathrm{FE}+U_\mathrm{rep}$ consists of the finitely extensible nonlinear elastic potential;
\begin{eqnarray}
U_\mathrm{FE}
&=& -\frac{C_\mathrm{FE}}{2} \sum_{i=1}^{N-1} (2a)^2 \log{\left(1-\frac{|\Vec{x}_{i+1}-\Vec{x}_i|^2}{(2a)^2} \right)}
\end{eqnarray}
and the repulsive potential between different beads;
\begin{eqnarray}
U_\mathrm{rep}
&=& \epsilon \sum_{i<j} \frac{a^{12}}{|\Vec{x}_i-\Vec{x}_j|^{12}}
\qquad \mathrm{for} \qquad |\Vec{x}_i-\Vec{x}_j| \leq R
\nonumber \\
&=& 0 \qquad \mathrm{for} \qquad |\Vec{x}_i-\Vec{x}_j| >R.
\nonumber
\end{eqnarray}
We set $\epsilon=k_BT$, $C_\mathrm{FE}=10k_BT/a^2$, $\gamma = (m k_BT/a^2)^{1/2}$ and the unit time is $\tau_{MD}=m/\gamma= (m a^2/k_BT)^{1/2} = \gamma a^2/k_BT$.
The total number of beads is $N=100$, for which the equilibrium size is $R\simeq 6.8a$ in 3D and $\simeq 9.5a$ in 2D.
Initial conditions are picked up from equilibrium configurations, and the force applied at $N$-th bead is switched on at $t=0$. We choose the time step for integrating equation of motion as $\delta t = 0.005 \tau_{MD} $.

Figure~2 (a) shows the time evolution of AD $\langle x(t) \rangle$ and the VD $\langle \Delta x^2(t) \rangle$ of the pulled bead by the force $f=0.3 k_BT/a$ in 3D (left) and 2D (right). Both quantities exhibit a slope close to $0.5$ in agreement with scaling predictions~(\ref{r*_f}) and~(\ref{DX_f}). After long time $t/\tau_{MD} > 5\times 10^3 \ (10^4)$ in 3D (2D), the dynamics becomes normal, i.e., $\langle x(t) \rangle \sim \langle \Delta x^2(t) \rangle \sim t$, which corresponds to the center of mass mode. This time is interpreted as the tension propagation time $\tau_f$ Eq.~(\ref{tau_f}), at which the effect of the driving force reaches to the other side of the chain end, and the chain settles in the steady-state as a whole. 

Figure~2 (b) shows the force dependence of the drift $\langle x(t) \rangle$. The data is taken at time $t/\tau_{MD} = 5 \times 10^2$ in the anomalous dynamics regime (see Fig.~2 (a)). Recalling the dynamical exponent $z=2+\nu^{-1}$ in the free draining case, the results are in good agreement with the predicted force exponent from Eq.~(\ref{r*_f}), that is $(2\nu)^{-1} \simeq 0.85 \ (0.67)$ in 3D (2D).

Figure~2 (c) shows the force dependence of the VD $\langle \Delta x^2(t) \rangle$. Again, the data is taken at time $t/\tau_{MD} = 5 \times 10^2$. The predicted force exponent from Eq.~(\ref{DX_f}) is $(2\nu)^{-1} -1 \simeq -0.15 \ (-0.33)$ in 3D (2D).
We find non-negligible deviations from the conjecture~(\ref{DX_f}), which is more evident in 2D.

Next, Fig.~3 shows the normalized VD-AD ratio $f\left< \Delta x^2(t) \right>/(k_BT\left< x(t)\right>)$.
Recall that the GLE with the relation~(\ref{FDT_GLE_f}) predicts this ratio to be time independent and exactly $2$ (eq.~(\ref{x_xx})).
While in 3D, this relation is satisfied reasonably well over a wide time range, appreciable deviations are found in 2D, which gets larger as the force becomes larger. 

The above results on the fluctuation $ \Delta x(t) $ of the tagged segment around the average drift suggests that, against a naive expectation, it cannot be described as the fBm in the strong force regime.
What causes it? To answer this, we shall take a microscopic polymer model, and attempt to derive the memory kernel and GLE based on the analysis of modes in the polymer chain.

\section{Mode Analysis}
\label{ModeAnalysis}
We first provide exact calculation for the Rouse model and deliver some remarks.
We then approximately incorporate the nonlinearity due to SA and HIs into the mode equation by making the effective friction and spring constants mode-number dependent (Sec.~\ref{SA_HI}).
Based on the solid framework on the weak force regime, we develop an approximate treatment in the strong force regime in Sec.~\ref{DrivenDynamics}.

\subsection{Rouse model}
\label{Rouse_model}
The polymer is modeled by $N+1$ connected beads. These beads have no excluded volume, linked by harmonic springs in series (the root-mean-square length of each spring is $a$), and move in a viscous fluid by being kicked by thermal noise and external force. As in Sec.~\ref{scaling_theory}, we again focus on the direction of the force only, and suppress the vector notations. In the limit where the bead labeling index $n \in [0,N]$ is made a continuous variable, the equation of motion reads
\begin{eqnarray}
\gamma \frac{\partial x_n}{\partial t} = k \frac{\partial^2 x_n}{\partial n^2} + \zeta_n + f_n,
\label{Rouse_eq}
\end{eqnarray}
where the friction coefficient for a bead $\gamma$ and the spring constant $k = 3 k_BT/a^2$ (in  3D) defines a microscopic time scale $\tau_0 = \gamma/k$, and $f_n$ is the external force acting on $n$-th bead.
Open boundary conditions are imposed at both chain ends for linear polymers
\begin{eqnarray}
\frac{\partial x_n}{\partial n} \Big|_{n=0}=\frac{\partial x_n}{\partial n} \Big|_{n=N}=0.
\label{Rouse_bound}
\end{eqnarray}
The random forces $\zeta_n(t)$ acting independently on individual beads are Gaussian white noise with zero mean, whose correlation obeys the FDT;
\begin{eqnarray}
\langle  \zeta_{n}(t)  \zeta_{m}(s) \rangle = 2 \gamma k_BT \delta(n-m)  \delta(t-s).
\label{FDT_segment}
\end{eqnarray}  

In the Rouse model, the noise and the external force acting on some segment affect the motion of other segments through the elastic connectivity. We may thus expect that the above FDT ~(\ref{FDT_segment}) at the individual segment level could be coordinated to generate a relation between the fluctuation and the response at the level of collective dynamics of the entire chain.
Below, we shall derive such a relation
\begin{eqnarray}
\langle \eta_n (t) \eta_m (s) \rangle =k_B T \chi_{nm}(t,s)
\label{FDT_nm}
\end{eqnarray}
with the concrete functional form of $\chi_{nm}(t,s)$ and $\eta_n(t)$. Here the response function $\chi_{nm}(t,s) \equiv \delta \langle {\dot x}_{n}(t) \rangle/ \delta f_{m}(s)$ describes the change in the average velocity of $n$-th segment at time $t$ caused by the force that acted on $m$-th segment at time $s (\le t)$, and $\eta_n(t)$ is a correlated Gaussian noise with zero mean $\langle \eta_n(t) \rangle =0$ acting on $n$-th segment. The FDT~(\ref{FDT_nm}) thus indicates that the cross correlation of the noise has a long time memory, which is related to the collective response of the segment. It includes Eq.~(\ref{FDT_GLE}) as a special case of the self-response and correlation $n=m$.

To this end, we analyze the Rouse equation~(\ref{Rouse_eq}) in terms of the normal coordinate $X_p(t)$
\begin{eqnarray}
X_p(t) = \int_{0}^{N} dn \  h_{p,n} x_n(t)
\end{eqnarray}
with
\begin{eqnarray}
h_{p,n}= \frac{1}{N}\cos{\left(\frac{ \pi pn}{N}\right)}.
\end{eqnarray}
Its inverse transform is
\begin{eqnarray}
x_n(t)&=&\sum_{p \ge 0} X_p(t) h^{\dagger}_{p,n}, 
\label{X_x}
\end{eqnarray}
where 
\begin{eqnarray}
h^{\dagger}_{p,n} = \frac{2 \cos{\left( \frac{\pi np}{N}\right)}}{1 + \delta_{p0}}.
\end{eqnarray}

The normal modes obey the following equation of the overdamped harmonic oscillator type
\begin{eqnarray}
\gamma_p \frac{\partial X_p(t)}{\partial t} = -k_p X_p(t) + Z_p(t) + F_p(t),
\label{Eq_NC}
\end{eqnarray}
where the spring and the friction constants $k_p$, $\gamma_p$ define the relaxation rate of the $p$-th mode $k_p/\gamma_p = (k/\gamma)(\pi p/N)^2$, and $Z_p = \int dn\  (\gamma_p/ \gamma) h_{p,n} \zeta_n(t)$ is the noise in the mode space.
The external force can be arbitrary, but for our present purpose, we manipulate a particular segment (labeled by the index $m$), i.e.,  $f_n(t) = f_m(t) \delta (n-m)$, which is distributed in the mode space according to  $F_p (t)= \int dn\  (\gamma_p/ \gamma) h_{p,n} f_n(t) =(\gamma_p/\gamma) h_{p, m} f_m(t)$.
There is no restoring force for $p=0$ mode, which corresponds to the motion of the center of mass. 

 It is useful to set as $\gamma_p = 2N \gamma/(1+ \delta_{p0})$,
thus $k_p = 2k ( \pi p)^2/N$, so that the FDT in the mode space takes a familiar form: $\langle Z_{p}(t) Z_{q}(s) \rangle = 2 \gamma_p k_BT  \delta_{pq}\delta(t-s)$. 
Equation~(\ref{Eq_NC}) is solved as
\begin{eqnarray}
 X_p(t) &=& \frac{1}{\gamma_p} \int_{t_0}^{t} ds \  e^{-(k_p/\gamma_p)(t-s)}(Z_p(s) + F_p(s)) \nonumber \\
 &&+ X_p(t_0)e^{-(k_p/\gamma_p)(t-t_0)},
 \label{Z_p_solution}
\end{eqnarray}
where $X_p(t_0)$ is the initial condition for the mode $p$.
We assume $t_0 \rightarrow - \infty$ so that the system is in a stationary state before we apply the external force.
By direct calculation, one can check the following FDT (see Appendix B)
\begin{eqnarray}
C_p(t,s) = k_BT \chi_p(t,s), 
\label{FDT_p}
\end{eqnarray}
where the velocity correlation function $C_p(t,s)$ and the response function $\chi(t,s)$ are defined as $ \langle {\Delta \dot X}_{p}(t) {\Delta \dot X}_{q}(s)\rangle \equiv \delta_{pq}C_p(t,s)$  and $\chi_p(t,s) \equiv \delta \langle {\dot X}_{p}(t) \rangle / \delta F_{p}(s)$, respectively.

Upon time derivative of Eq.~(\ref{Z_p_solution}) and transforming it into the real coordinate using Eq.~(\ref{X_x}), one can express the time derivative of the position of $n$-th segment in the following form;
\begin{eqnarray}
\frac{dx_n(t)}{dt} = \int_{-\infty}^t ds \ \chi_{nm}(t,s) f_m(s) + \eta_n(t),
\label{GLE}
\end{eqnarray}
where 
\begin{eqnarray}
\chi_{nm}(t,s) = \sum_{p \ge 0} \chi_p (t,s) h^{\dagger}_{p,n}h^{\dagger}_{p,m} 
\label{chi_v}
\end{eqnarray}
and
\begin{eqnarray}
\eta_n(t)= \sum_{p \ge 0}\int_{-\infty}^t ds \ \chi_p(t,s) Z_p(s) \ h_{p,n}^{\dagger} 
\end{eqnarray}
are, respectively, interpreted as the velocity response function and the noise (The functional form of $\chi_p(t,s)$ is given in Eq.~(\ref{chi_p})). The latter is identified as the velocity fluctuation $\eta_n(t) = \Delta {\dot x}_n(t) = {\dot x}_n(t) - \langle  {\dot x}_n(t) \rangle$ of $n$-th segment. The fluctuation-response relation~(\ref{FDT_nm}) can be most easily verified by decomposing the response and the correlation functions into modes:
Eq.~(\ref{chi_v}) and $\langle \eta_n (t) \eta_m (s) \rangle =\langle \Delta {\dot x}_n(t) \Delta {\dot x}_m(s)\rangle=\sum_{p \ge 0} C_p(t,s) h^{\dagger}_{p,n}h^{\dagger}_{p,m}$. The FDT~(\ref{FDT_p}) in the mode space then leads to Eq.~(\ref{FDT_nm}).

{\it Memory kernel---.}
Suppose we tag a particular segment (labeled by $m$). The force $f(t)$ is applied only to that $m$-th segment, and we track its stochastic motion. The information of interest is contained in the self-response $\chi_{mm}(t,s)$. In this context of the single segment tracking analysis, positions of other segments except for $m$ are inaccessible to observations. One can therefore omit the label index and write the equation of motion of the tagged segment as
\begin{eqnarray}
\frac{dx(t)}{dt} = \int_{-\infty}^t ds \ \mu(t-s) f(s) + \eta(t)
\label{Eq_tagged_monomer}
\end{eqnarray}
where, as verified above, the FDT~(\ref{FDT_GLE}) holds. The mobility kernel $\mu(t) (\equiv \chi_{mm} (t,0))= \mu_0^{(CM)}(t) + \mu_0(t) + \mu_M(t)$ composes of three terms according to  Eq.~(\ref{chi_v});
\begin{eqnarray}
\mu_0^{(CM)} (t) &=& \frac{2}{\gamma_0} \delta(t) = \frac{2}{N \gamma}\delta(t) \label{mu_CM}\\
\mu_0(t) &=&  \sum_{p =1}^N \frac{2}{\gamma_p} \delta(t) (h_{p,m}^{\dagger})^2 \simeq \frac{2}{\gamma} \delta(t) \label{mu_m0}\\
\mu_M(t) &=& - \sum_{p =1}^N \frac{k_p}{\gamma_p^2}e^{-(k_p/\gamma_p)t}(h_{p,m}^{\dagger})^2 \label{mu_m_Rouse} \\
&\simeq& -\frac{1}{4 \sqrt{\pi}  }\frac{1}{\gamma \tau_0} \left( \frac{t}{\tau_0}\right)^{-3/2} \quad (\tau_0 \ll t \ll \tau), 
\nonumber
\end{eqnarray}
where $\tau \equiv \gamma_1/k_1 \simeq \tau_0 N^2$ is the terminal (Rouse) time. In the above, the upper bound in the summation over $p$ is set reflecting the original discrete nature of the model with $N$ degrees of freedom, and we replace $\cos^2{(p\pi n/N)}$ by the average $1/2$. The last near-equality in Eq.~(\ref{mu_m_Rouse}) is valid in the time window $\tau_0 \ll t \ll \tau$, where the summation over $p$ can be replaced by the Gaussian integral. For longer time $t > \tau$, only the $p=1$ mode is relevant, thus, the memory decays exponentially.

{\it Remarks---.} 

(i) In the very short time scale ($ t \simeq \tau_0$), the segment is unaware of the connectivity, and exhibits a usual viscous response with the segment friction coefficient $\gamma$ (Eq.~(\ref{mu_m0})). In the time scale coarser than $\tau$, the response again becomes viscous, but now with the much larger friction coefficient $\gamma N$, i.e., the center of mass mode (Eq.~(\ref{mu_CM})); see the sum rule Eq.~(\ref{sum_rule}) below.

(ii) In the intermediate time scale ($\tau_0 \ll t \ll \tau$), the last term $\mu_M(t)$ (Eq.~(\ref{mu_m_Rouse})) dominates the dynamics of the tagged segment. Noting $z=4$ for a Rouse model, the result~(\ref{mu_m_Rouse}) agrees with the scaling analysis Eq.~(\ref{mu_eq}) in Sec.~(\ref{scaling_theory}). This term represents a memory effect, which arises from the superposition of the internal modes in the Rouse polymer. The minus sign here is a hallmark of the viscoelastic response inherent in the system with elastic connectivity (see, for instance, a similar analysis for polymerized membrane~\cite{Mizuochi}).

(iii)
In Eq.~(\ref{mu_m_Rouse}), the mode
\begin{eqnarray}
p^*(t) = \left( \frac{\gamma N^2}{k \pi^2 t}\right)^{1/2} \simeq \left( \frac{t}{\tau}\right)^{-1/2}
\label{p*_eq}
\end{eqnarray}
has the largest contribution at time $t$. The corresponding number of segments is $n^{*}(t) \simeq N/p^*(t) \simeq (t/\tau_0)^{1/2}$. This agrees with our scaling estimate for the tension front in the weak force regime (Eq.~(\ref{n*_eq}) with $z=4$ and $\nu=1/2$).
The physics behind this agreement is the following;
The effect from the larger scale beyond the tension front $n^*(t)$ is irrelevant, or only marginal. Therefore, even if we neglect it, i.e., by shifting the lower bound $p_{l.b}$ in the summation $p_{l.b}=1$ to $p_{l.b}= p^*(t)$, one should get a qualitatively correct result with the proper exponent.

(iv) The following sum rule 
\begin{eqnarray}
\int_0^{\infty} dt \  [\mu_0(t) + \mu_M(t)] =0
\label{sum_rule}
\end{eqnarray}
may hold for any physical system, which indicates that in the long time limit $t \gg \tau$, all the internal modes relax, and only the center of mass mode remains.

(v) In certain visco-elastic environments, the segment response itself could contain the memory effect. Then, one may think of the visco-elastic Rouse model, where the viscous friction term $\gamma {\dot x}_n$ in the Rouse equation (\ref{Rouse_eq}) is replaced with the integral kernel $\int ds \ \Gamma_1(t-s) {\dot x}_n(s)$~\cite{PRL_Weber_2010, PRE_Weber_2010,Rouse_viscoelastic}. In the case of the power-low memory kernel $\Gamma_1(t) \sim t^{-\alpha_1}$ with $0< \alpha_1 < 1$, the exponential relaxation in viscous Rouse model (Eq.~(\ref{Z_p_solution})) is generalized to the non-exponential one described by the generalized Mittag-Leffler function. This results in the memory kernel in the tagged segment dynamics $\mu_M(t) \sim - t^{-(2-(\alpha_1/2))}$, hence the anomalous exponent $\alpha = \alpha_1/2$ for the tagged segment diffusion. Such a visco-elastic Rouse model has been proposed to analyze the sub-diffusion of bacterial chromosomal loci~\cite{BJ_Lampo_2015,PRL_Weber_2010}. The usual viscous result corresponds to the limit $\alpha_1 \rightarrow 1$. The relation between exponents for the single segment exponent $\alpha_1$ and the tagged segment one $\alpha$ with a factor $2$ is a general consequence of the Rouse model. 

(vi) The Rouse model is valid as long as $f \lesssim k_BT/a$. For larger force, the chain section close to the pulled site becomes highly stretched, forming a ``stem"~\cite{EPL_Brochard_1995}. For the prescription and the scaling analysis in such a situation, see Ref.~\cite{PRE_Sakaue_2012}.

\subsection{Self-avoidance and hydrodynamic interactions}
\label{SA_HI}
In many of practical situations, one of or both of these effects become important. These interactions are non-local, and conformation dependent, hence, make the equations of motion highly nonlinear, which prevent the rigorous analysis based on the mode expansion. Nevertheless, one can gain physically appearing insights in terms of approximate mode analysis. The pre-averaging approximation provides a way to treat the HIs in terms of modes, in which the conformation-dependent mobility tensor is averaged using the equilibrium segment distribution~\cite{Doi_Edwards}. This yields for the effective friction constant for the mode $p$
\begin{eqnarray}
\gamma_p &\simeq& \left\{
\begin{array}{ll}
\gamma N^{\nu(z-2)} & (p=0)  \\
\gamma p (N/p)^{\nu(z-2)} & (p \neq 0) \label{gamma_p} 
\end{array}
\right.
\end{eqnarray}
Note that the free-draining polymer $z=2 + (1/\nu)$ bears no relation to the pre-averaging; we then recover $\gamma_p$ for the Rouse model.

In a similar level, the self-avoidance (SA) can be treated by employing the linearization (Gaussian) approximation, which alters the spring constant for the mode $p$ as
\begin{eqnarray}
k_p &\simeq& k p(p/N)^{2 \nu}. \label{k_p}
\end{eqnarray}
The validity of this form as well as a high degree of statistical independence of different modes has been numerically demonstrated in Ref.~\cite{Panja_Rouse_mode}.
Note that for the ideal polymer $\nu = 1/2$, we recover $k_p$ for the Rouse model.

Let us analyze the mode equation~(\ref{Eq_NC}) with Eqs.~(\ref{gamma_p}) and~(\ref{k_p}). Note that the terminal time is now given by $\tau = \gamma_1/k_1 \simeq  \tau_0 N^{\nu z}$ (see Eq.~(\ref{tau_eq})).
The mobility kernel is again decomposed as  $\mu(t) = \mu_0^{(CM)}(t) + \mu_0(t) + \mu_M(t)$. While the segment instant response $\mu_0(t)$ is essentially unchanged from the Rouse model (Eq.~(\ref{mu_m0})), the $N$ dependence of the center of mass response is modified as $\mu_0^{(CM)} (t) \simeq (\gamma N^{\nu(z-2)})^{-1} \delta(t)$. In addition, the memory kernel is evaluated as
\begin{eqnarray}
\mu_M(t) &\simeq&  - \sum_{p=1}^{N} \frac{1}{\gamma N \tau_0} \left( \frac{p}{N}\right)^{2\nu(z-1)-1}e^{-\frac{t}{\tau_0}\left( \frac{p}{N}\right)^{\nu z}}  (h_{p,n}^{\dagger})^2 \nonumber \\
&\simeq&   - \frac{1}{\gamma \tau_0} \left( \frac{t}{\tau_0}\right)^{-(2-2z^{-1})} \label{mu_m_} \quad (\tau_0 \ll t \ll \tau) \label{mu_M_eq}
\end{eqnarray}
The last near-equality is valid in the intermediate time scale, where the summation is evaluated as the integral using the formula\footnote{The symbol $\Gamma(\cdot)$ here is used as the gamma function $\Gamma(z)=\int_0^{\infty} u^{z-1}e^{-u} du$ in this integral formula.} $\int_0^{\infty}dx \ x^{b-1}e^{-a x^{\theta}} = \Gamma(b/\theta) /(\theta a^{b/\theta})$ for $a, b, \theta > 0$.
The result agrees with our scaling argument (see Eq.~(\ref{mu_eq})), and the tagged segment dynamics in this time scale is a fBm with the anomalous exponent $\alpha = 2/z$.

{\it Remarks---.}

(i) At time $t$, the mode $p^*(t)  \simeq \left( t/\tau \right)^{-1/(\nu z)}$
has the largest contribution. The corresponding number of segments $n^*(t) \simeq N/p^*(t)$ agrees with our scaling argument Eq.~(\ref{n*_eq}) for the dynamics of tension front.

(ii) For the present description to be valid, at least qualitatively, the condition $f \lesssim k_BT/(aN^{\nu})$ is required. For stronger force, the conformation of the polymer is markedly deviated from the equilibrium distribution, which invalidates the assumption used to evaluate effective friction and spring constants in Eqs.~(\ref{gamma_p}) and~(\ref{k_p}). This is contrasted to the Rouse model case, for which the condition is much weaker, associated to the bond stretching (See remark (vi) in Sec.~\ref{Rouse_model}), but not the conformation.

In Sec.~\ref{DrivenDynamics}, we aim at constructing an effective description, which allows us to analyze the fluctuating driven dynamics in larger force regime $f \gtrsim k_BT/(aN^{\nu})$ even with SA and HIs.

\subsection{Driven dynamics}
\label{DrivenDynamics}
Suppose we start applying a constant strong force $f > k_BT/(aN^{\nu})$ to the chain end ($N$-th segment) at time $t=0$ (see Fig.~1(b)). Before that moment ($t<0$), the polymer assumes an equilibrium conformation. We are interested in the motion of that pulled segment. The dynamics is nonlinear with SA or HIs and nonequilibrium in strong force regime. To analyze the average dynamics, the following nonlinear diffusion-type equation (called {\it p}-Laplacian diffusion equation) has been proposed~\cite{EPL_Brochard_1994,EPL_Sebastian_2011}:
\begin{eqnarray}
\frac{\partial x_n}{\partial t} = D_0 \frac{\partial}{\partial (na)}\left[\left( \frac{\partial x_n}{\partial (na)}\right)^{p-2} \left( \frac{\partial x_n}{\partial (na)}\right) \right]
\label{p_Lap}
\end{eqnarray}
with $p=(z-2)\nu/(1-\nu)$ and the segment diffusion coefficient $D_0 \simeq k_BT/\gamma$.
This equation is derived based on the force balance argument for the chain of blobs, and can be thought of as a nonlinear extension of Rouse model (see Appendix A for the derivation).

Equation~(\ref{p_Lap}) would be a useful starting point to analyze the stochastic dynamics of  the polymer stretching in terms of the collective mode in the chain. Since, by construction, this is expected to provide a reasonable description on the average dynamics, one may add a random force $\zeta_n$ of zero mean to get a nonlinear Langevin equation
\begin{eqnarray}
\gamma_n \frac{\partial x_n}{\partial t}=k_n \frac{\partial^2 x_n}{\partial n^2}  +\zeta_n(t) +f_n,
\end{eqnarray}
where $k_n$ and $\gamma_n$ are given by Eqs.~(\ref{k_n_s}) and~(\ref{gamma_n_s}) in Appendix~A. Within the mono-block approximation, one can linearize it
\begin{eqnarray}
\gamma^{(f)} \frac{\partial x_n}{\partial t} = k^{(f)} \frac{\partial^2 x_n}{\partial n^2} + \zeta_n^{(f)}(t) + f_n
\label{f_Rouse}
\end{eqnarray}
with the force-dependent spring and friction coefficients Eqs.~(\ref{k_f}),\ (\ref{gamma_f}).
The random forces $\zeta_n$ independently acting on individual segments are assumed to be Gaussian white noise with zero mean $\langle \zeta_n^{(f)}(t) \rangle =0$, whose correlation obeys the FDT; $\langle \zeta_n^{(f)}(t)\zeta_m^{(f)}(s) \rangle = 2 \gamma^{(f)}k_BT \delta(n-m) \delta(t-s)$. 
Equation~(\ref{f_Rouse}) reduces to the Rouse model (eq.~(\ref{Rouse_eq})) when $\nu=1/2$ and $z=2 + (1/\nu) = 4$. Otherwise, the SA or HIs result in the nonlinear response. 
With the force free boundary condition (Eq.~(\ref{Rouse_bound})) and explicit inclusion of the external force $f_n(t)= f \Theta(t) \delta (N-n)$ acting on the end segment $n=N$, one can follow the analysis developed in Secs.~\ref{Rouse_model} -~\ref{SA_HI}.

In analyzing the dynamics of nonlinear response, one has to be aware of the change in the mode spectrum, i.e., $(k_p,\gamma_p) \rightarrow (k_p^{(f)},\gamma_p^{(f)}) $ due to the external force.
This effect may be treated in the following way.
After the force is switched on at $t=0$, the equilibrium mode would persist during the induction time (see remark (i) in Sec.~\ref{time_dependent_friction}). Equation~(\ref{Eq_NC}) with Eqs.~(\ref{gamma_p}) and~(\ref{k_p}) would be thus valid up to $t=\tau_{f0}$. At $t > \tau_{f0}$, the effect of the driving dominates the mode dynamics, and the spring and the friction constants become altered to those in the strong force regime. Thus, making these constants time dependent, the equation of motion in mode space becomes
\begin{eqnarray}
\gamma^{*}_p (t) \frac{\partial X_p(t)}{\partial t} = -k^{*}_p (t) X_p(t) + Z^{*}_p (t) + F_p (t),
\label{Eq_NC_f}
\end{eqnarray}
where the equilibrium mode structure $\gamma^{*}_p (t) = \gamma_p$, $k^{*}_p (t) = k_p$, $Z^{*}_p (t) = Z_p(t)$ persists only up to $t < \tau_{f0}$ (Eqs.~(\ref{gamma_p}) and~(\ref{k_p})). At  $t > \tau_{f0}$, these switch to the stretched mode  $\gamma^{*}_p (t) = \gamma^{(f)}_p$, $k^{*}_p (t) = k^{(f)}_p$, $Z^{*}_p (t) = Z^{(f)}_p(t)$ with
\begin{eqnarray}
\gamma^{(f)}_p &\simeq& \left\{
\begin{array}{ll}
N \gamma^{(f)} & (p < p_f)  \\
\gamma p (N/p)^{\nu(z-2)} & (p >p_f) \label{gamma_p_f} 
\end{array}
\right.
\end{eqnarray}
\begin{eqnarray}
k^{(f)}_p &\simeq& \left\{
\begin{array}{ll}
k^{(f)}  p^2/N & (p < p_f)  \\
k p(p/N)^{2 \nu} & (p >p_f) \label{k_p_f} 
\end{array}
\right. 
\end{eqnarray}
and the Gaussian white noise $Z_p^{(f)}(t)$ with zero mean and the correlation $\langle Z^{(f)}_{p}(t) Z^{(f)}_{q}(s) \rangle = 2 \gamma^{(f)}_p k_BT  \delta_{pq}\delta(t-s)$.
The tension propagation time given in Eq.~(\ref{tau_f}) can be identified as the slowest relaxation time in the stretched mode $\tau_f = \gamma_1^{(f)}/k_1^{(f)}$.
Here, the characteristic mode number is introduced 
\begin{eqnarray}
p_f = 2N \left( \frac{fa}{k_BT}\right)^{1/\nu} \simeq \frac{N}{g}
\label{p_f}
\end{eqnarray}
in such a way that the effect of the force is negligible for modes with $p>p_f  (\Leftrightarrow a (2N/p)^{\nu} < k_BT/f)$, hence, the friction and spring constants are given by those in weak force regime (Eqs.~(\ref{gamma_p}) and~(\ref{k_p}), respectively). Notice that our construction assures continuity for both $\gamma_p^{(p)}$ and $k_p^{(f)}$ across $p_f$.

The solution of Eq.~(\ref{Eq_NC_f}) for $t>\tau_{f0}$ takes the same form as Eq.~(\ref{Z_p_solution}) with the replacement $(k_p,\gamma_p, t_0) \rightarrow (k_p^{(f)},\gamma_p^{(f)}, \tau_{f0}) $, where the ``initial" condition is given by
\begin{eqnarray}
X_p(\tau_{f0}) &=& \frac{1}{\gamma_p} \int_{-\infty}^{\tau_{f0}} ds \  e^{-\frac{k_p}{\gamma_p}(\tau_{f0}-s)}Z_p(s) 
\nonumber \\
&& + \frac{F_p}{\gamma_p}\left( 1- e^{-\frac{k_p}{\gamma_p} \tau_{f0}}\right).
\label{Xp_ini}
\end{eqnarray}
Crucially, this noise from the ``initial" condition adds an extra contribution to the noise $\eta_f(t)$ in the motion of the tagged segment, which leads to deviation from the relation~(\ref{FDT_GLE_f}) due to the SA effect as will be discussed below.

The motion of the tagged segment is described by Eq.~(\ref{Eq_tagged_monomer}) in the weak force regime at $t<\tau_{f0}$. To take account of the changes in spring and friction constants at $t > \tau_{f0}$, we need to modify it as
\begin{eqnarray}
\frac{dx(t)}{dt} = \int_{-\infty}^t ds \ \mu_f(t,s; \tau_{f0}) f(s) + \eta_f(t; \tau_{f0})
\label{Eq_tagged_monomer_f}
\end{eqnarray}
with the mobility kernel
\begin{eqnarray}
\mu_f(t,s;\tau_{f0}) =  \sum_{p \ge 0} \chi_{p,f} (t,s;\tau_{f0}) (h^{\dagger}_{p,N})^2 
\label{mu_f}
\end{eqnarray}
and the noise
\begin{eqnarray}
\eta_f(t; \tau_{f0}) &=&\sum_{p \ge 0} h_{p,N}^{\dagger}\Bigl[ \int_0^t ds \chi_{p,f}(t,s;\tau_{f0}) Z_p^{*}(s) \nonumber \\
&&- X_p(0) \frac{k_p^{*}(t)}{\gamma_p^{*}(t)}\exp{\left(- \int_0^t ds \frac{k_p^{*}(s)}{\gamma_p^{*}(s)} \right)} \Bigr]
\label{eta_f}
\end{eqnarray}
where $\chi_{p, f}(t,s;\tau_{f0})$ is the velocity response function for the mode $p$ calculated in Appendix B.

\begin{figure}[t]
\begin{center}
\includegraphics[scale=0.5]{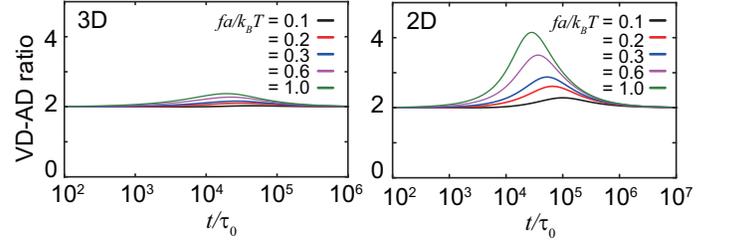}
      \caption{
	  (Color Online) 
	  VD-AD ratio $f \left<\Delta x^2 \right> / (k_BT\left< x \right>) $ obtained from theory (Eq.~(\ref{x_xx_diff})) in 3D (left) and 2D (right) as a function of time. 
	  	  The calculation was carried out with $N=100$.
		  Note that there are uncertainties in the precise values in the peak heights due to the scaling estimate of spring constants. In the above plots, we set all the numerical constants (including $(h_{p,N}^\dagger)^2$) to be unity.
	  }
\label{fig4}
\end{center}
\end{figure}

{\it Memory kernel---.}
The mobility kernel $\mu_f(t,s; \tau_{f0})$ in the intermediate time scale $\tau_{f0} \ll s < t \ll \tau_f$ is stationary, and dominated by the memory effect due to the connectivity, that is $\mu_f(t,s; \tau_{f0}) \simeq \mu_{M,f}(t-s)$ with
\begin{eqnarray}
&&\mu_{M,f}(t) 
\simeq - \sum_{p \ge 1 } \frac{k_p^{(f)}}{(\gamma_p^{(f)})^2}e^{-\frac{k_p^{(f)}}{\gamma_p^{(f)}}t}(h_{p,N}^{\dagger})^2 \nonumber \\
&= & 
 - \sum_{p=1}^{p_f} \frac{1}{\gamma N \tau_{f0}} \left( \frac{p}{p_f}\right)^{2} \left( \frac{fa}{k_BT}\right)^{z-2- \nu^{-1}} e^{-\frac{t}{\tau_{f0}}\left( \frac{p}{p_f}\right)^{2}}  (h_{p,N}^{\dagger})^2 \nonumber \\
 &&-\sum_{p=p_f}^{N} \frac{1}{\gamma N \tau_0} \left( \frac{p}{N}\right)^{2\nu(z-1)-1}e^{-\frac{t}{\tau_0}\left( \frac{p}{N}\right)^{\nu z}}  (h_{p,N}^{\dagger})^2 
\label{mu_m_f}
\end{eqnarray}
In the time window $\tau_{f0} \ll t \ll \tau_f$, the second term is negligible and the summation in the first term can be approximated by the Gaussian integral. This calculation leads to our scaling estimate~(\ref{mu_f_scaling}) in Sec.~\ref{scaling_theory}. For longer time $t > \tau_f$, the memory decays exponentially, and only the center of mass ($p=0$) mode remains. It is characterized by the viscous response with the friction coefficient $\simeq N \gamma^{(f)}$.

{\it Fluctuation and response relation---.}
The correlation function of the velocity fluctuation is 
\begin{eqnarray}
\langle \eta_f(t;\tau_{f0}) \eta_f(s;\tau_{f0}) \rangle= \sum_{p \ge 0} C_{p,f}(t,s;\tau_{f0})  (h_{p,N}^{\dagger})^2,
\end{eqnarray}
where $C_{p,f}(t,s) $ is the correlation function in the mode space calculated in Appendix B.
Comparing this with Eq.~(\ref{mu_f}), we find
\begin{eqnarray}
 &&\langle \eta_f(t;\tau_{f0}) \eta_f(s;\tau_{f0}) \rangle - k_BT \mu_f(t,s;\tau_{f0}) \nonumber \\
&&= \left\{
\begin{array}{lll}
0 & (s<t < \tau_{f0})  \\
0 & (s< \tau_{f0} < t)  \\
  \sum_{p \ge 1}   C_{p,f}^{(ex)}(t,s;\tau_{f0}) (h_{p,N}^{\dagger})^2 & ( \tau_{f0}<s<t)
\label{C-mu} 
\end{array}
\right. 
\end{eqnarray}
Here, a factor from the ``initial" condition can be evaluated using Eq.~(\ref{X_p_equilibrium}). This leads to ``zero" in the first and second lines, and the expression of the excess term in the third line in terms of the change in the spring constant
\begin{eqnarray}
C_{p,f}^{(ex)}(t,s;\tau_{f0}) 
&=& 
\left(\frac{k_p^{(f)}}{\gamma_p^{(f)}}\right)^2 \left( \frac{k_BT}{k_p} -\frac{k_BT}{k_p^{(f)}} \right)
\\
&& \times
 e^{-(k_p^{(f)} /\gamma_p^{(f)})(t+s-2\tau_{f0})} 
\nonumber
\end{eqnarray}
This term breaks the time translational invariance, and decays exponentially with a characteristic rate $k_p^{(f)}/\gamma_p^{(f)}$.
From this, we obtain\footnote{Begining with the solution $X_p(t)$ of Eq.~(\ref{Eq_NC_f}) makes the calculation easier to check Eq.~(\ref{x_xx_diff}).}
\begin{eqnarray}
&&\left< \Delta x(t)^2 \right>
-\frac{2k_BT}{f}\left< x(t) \right>
\label{x_xx_diff}
\\
&=& 
\sum_{p \ge 1}\left( \frac{k_BT}{k_p}-\frac{k_BT}{k_p^{(f)}} \right)
\left( 1-e^{-\frac{k_p^{(f)}}{\gamma_p^{(f)}}(t-\tau_{f0})} \right)^2(h_{p,N}^{\dagger})^2.  \nonumber
\end{eqnarray}
The deviation (right hand side in eq.~(\ref{x_xx_diff})) is positive due to the force induced hardening $k_p^{(f)} > k_p$ (for SA chain).
Comparing the 2D and 3D cases, the larger deviation is expected for 2D as $k_{p,\mathrm{2D}}^{(f)} > k_{p,\mathrm{3D}}^{(f)}$.
In addition, Eq.~(\ref{x_xx_diff}) tells that the deviation grows as time evolves until $\tau_{f}$.
This means that the VD-AD ratio $f \langle \Delta x^2(t) \rangle /(k_BT \langle x(t) \rangle)$ peaks around $\tau_{f}$. 
At $t > \tau_{f}$,  $p=0$ mode in the denominator overwhelms the internal modes $p\geq 1$, leading to the recovery of the relation $f\left< \Delta x^2\right>/\left< x\right>=2k_BT$.
These trends are clearly seen in Fig. 4, where we plot the VD-AD ratio obtained from the above theory. 
Our present treatment is rather crude in the sense that the switching to the strong force regime around $t \simeq \tau_{f0}$ is treated through a discrete jump in the effective parameters. In reality, it would take place more smoothly. Nevertheless, our theoretical prediction well captures the essential trend in the MD simulation results in Fig.~3.

{\it Remarks---.}

(i) At time $t$, the mode $p^*(t) \simeq (t/\tau_f)^{-1/2}$ has the largest contribution in Eq.~(\ref{mu_m_f}). This corresponds to the number of segments $n^*(t) \simeq N/p^*(t)$, which agrees with our scaling estimate~(\ref{n*_f}) for the tension front in the strong force regime.

(ii) The mode with $p < p^*(t)$ may be unphysical for the stretching process as such a large scale part is not stretched by the force yet.
However, we expect that these fictive modes do not alter the qualitative conclusion on the dynamics of tension propagation, just as the case in the weak force regime (see remark (iii) in Sec.~\ref{Rouse_model}). 

(iii) 
A rough estimate in the peak height in VD-AD ratio can be obtained by evaluating $p=1$ mode in Eq.~(\ref{x_xx_diff});
\begin{eqnarray}
\frac{f \left<\Delta x^2(\tau_{f}) \right> }{ k_BT\left< x (\tau_{f})\right> }&\sim& 2 + f \frac{k_1^{-1}- [k_1^{(f)}]^{-1}}{f\tau_f/\gamma_0^{(f)}}(1- e^{-1})^2 (h_{p,N}^{\dagger})^2 \nonumber \\
&\sim& 2 + \left[ \left( \frac{fa N^{\nu}}{k_BT}\right)^{(2\nu-1)/\nu} -1\right]  c_0
\end{eqnarray}
With a factor $c_0 \sim (1- e^{-1})^2(h_{p,N}^{\dagger})^2 \sim 0.4$, we obtain VD-AD ratio $ \sim 5.6$ and $\sim 2.4$ in 2D and 3D cases, respectively, with $N=100$ and $fa/k_BT = 1$.
Despite indefiniteness of these numerical values (see the caption of Fig.~4), this estimate would be useful to see the qualitative dependence on the force and the chain length.

(iv)
The tension propagation time $\tau_f$ fluctuates in each realization of the stretching processes, and Eq.~(20) is regarded as the average $\langle \tau_f \rangle = \gamma_1^{(f)}/k_1^{(f)}$. In the strong force regime, the dominant source of the stochasticity comes not from the noise $Z_p(t)$ but from the initial conformation of the polymer along which the tension propagates. In terms of the mode analysis, the fluctuation in $\tau_f$ can be evaluated in the following way.

For clarity of the argument, suppose that the force is strong enough $f \simeq k_BT/a$ so that the induction time is very short $\tau_{f0} \simeq \tau_0$. Neglecting the noise effect, the displacement of the center of mass ($p=0$) mode and the slowest relaxational ($p=1$) mode are, respectively, $X_0(t)-X_0(0) =F_0 t/\gamma_0^{(f)}$ and $X_1(t)-X_1(0) =(F_1/k_1^{(f)}-X_1(0))(1-e^{-k_1^{(f)}t/\gamma_1^{(f)}})$.
Comparing these, one finds that at $t= \langle \tau_f \rangle $, the displacement in $p=0$ mode reaches the final displacement in $p=1$ mode on average, i.e., $\langle \tau_f \rangle \simeq \gamma_0^{(f)} F_1 /(F_0 k_1^{(f)}) \simeq \gamma_1^{(f)}/k_1^{(f)}$.
Taking account of the fluctuation in the latter due to the initial distribution, one obtains the variance in the propagation time as $\langle(\Delta \tau_f)^2 \rangle \simeq ( \gamma_0^{(f)}/F_0 )^2 \langle X_1(0)^2 \rangle$. Evaluating the variance in the initial distribution using Eq.~(\ref{X_p_equilibrium}), this leads to
\begin{eqnarray}
\sqrt{\langle(\Delta \tau_f)^2 \rangle} \simeq \tau_0 N^{1+\nu}\left( \frac{fa}{k_BT}\right)^{1-z+(1/\nu)}
\end{eqnarray}
The same result has been obtained in ref.~\cite{PRE_Saito_2012} using the scaling argument.

(v) Events on the short length and time scales are described by the weak force regime. Such a range increases with the decrease in the force; the characteristic mode number and the time change from $p_f \simeq N$ and $\tau_{f0} \simeq \tau_0$ at $f\simeq k_BT/a$ to $p_f \simeq 1$ and $\tau_{f0} \simeq \tau$ at $f \simeq k_BT/R$. 

(vi) Our theory indicates that it is the nonlineality in the elastic response (force-dependent spring constant), but not the frictional response, that is responsible for the non-trivial VD-AD ratio in the stretching process. The sign of Eq.~(\ref{x_xx_diff}) depends on whether the system exhibits the stiffening or the softening under the force.

\section{Concluding Remarks}
It has been long known that a tagged segment in a polymer exhibits a sub-diffusion in the intermediate time scale, and its consequence ranges from the dynamical function of biopolymers to the rheology of polymer solutions.
In this paper, we formulated the problem in terms of the mobility problem, i.e., the dynamical response of the segment after the application of external force, and introduced the weak and strong force regimes for the anomalous dynamics.

In the weak force (equilibrium) regime, the motion in the intermediate time scale is dominated by the memory effect, leading to a conventional fBm.
We performed lucid and exact analysis for a Rouse model, which leads to a microscopic basis for the fBm. The deduced memory kernel fully agrees with a simple scaling argument based on the physical picture of tension transmission. Together with the approximation scheme to include the SA and HIs, we believe that the present approach provides a comprehensive picture on the anomalous dynamics of the tagged segment in the weak force regime.

In the strong force (driven) regime, the motion in the intermediate time scale is again dominated by the memory effect arising from the tension transmission, but now the tension dynamics accompanies a large conformational distortion, and qualitatively different from that in the weak force regime.
We discussed that the memory kernel generally becomes force dependent, from which one can derive the nonlinear dynamical scaling for the anomalous drift.
Unlike the weak force regime, the fluctuation and the response do not necessarily satisfy a simple proportionality relation due to the noise from the ``initial condition" at $t=\tau_{f0}$, after which the dynamics enters the strong force regime. This extra noise is non-stationary, making the fluctuating dynamics to deviate from the fBm.  
On the basis of the approximate mode analysis, we proposed a formula to relate the fluctuation and the response in the driven process, which is in a rather good agreement with results obtained from MD simulations.

A recent study has made use of the VD-AD ratio of the labeled locus of bacterial chromosome during the segregation process to estimate its driving force~\cite{BJ_Lampo_2015}. We feel that our present study could be a useful guide for such an analysis.

\section*{Acknowledgement}
This work was supported by KAKENHI [Grant No.26103525,``Fluctuation and Structure", Grant No.24340100, Grant-in-Aid for Scientific Research (B)], Ministry of Education, Culture, Sports, Science and Technology (MEXT), Japan and JSPS Core-to-Core Program (Nonequilibrium Dynamics of Soft Matter and Information).

\section*{Appendix A}
\label{deterministic}
We review the previous works on the deterministic (average) dynamics. 

{\it Coarse-grained description: a chain of blobs ---.}

To discuss the dynamics in the scale larger than the blob size, we envision the stretched polymer as a chain of blobs. The blobs are labeled with the index ${\tilde n} = 0,1,2, \cdots$ from the free end at the rear. The ${\tilde n}$-th blob comprised of $g_{{\tilde n}}$ segments has the spatial size $\xi_{{\tilde n}} \simeq a g_{{\tilde n}}^{\nu}$. By taking the $x$ axis as the pulling direction, the position ${\tilde x}_{{\tilde n}}$ of the center of ${\tilde n}$-th blob is
\begin{eqnarray}
 {\tilde x}_{{\tilde n}}= {\tilde x}_{{\tilde n}=0}+ \sum_{{\tilde n}=0}^{{\tilde n}-1}\xi_{{\tilde n}}.
 \label{Eq.1}
 \end{eqnarray}
 The dynamics of the chain of blobs can be analyzed by noting that the spring and the friction constants ${\tilde k}_{{\tilde n}}$, ${\tilde \gamma}_{{\tilde n}}$ of the ${\tilde n}$-th blob are given by
 \begin{eqnarray}
 {\tilde k}_{{\tilde n}} &\simeq& \frac{k_BT}{\xi_{{\tilde n}}^2} \label{K_n_blob}\\
 {\tilde \gamma}_{{\tilde n}} &\simeq& \gamma \left( \frac{\xi_{{\tilde n}}}{a}\right)^{z-2},
 \end{eqnarray}
which lead to the force balance equation
\begin{eqnarray}
{\tilde f}_{{\tilde n}}^{(el)} + {\tilde f}_{{\tilde n}}^{(vis)}=0
\label{f_balance}
\end{eqnarray}
with the elastic restoring force
\begin{eqnarray}
{\tilde f}_{{\tilde n}}^{(el)}&=&{\tilde k}_{{\tilde n}}({\tilde x}_{{\tilde n}+1} - {\tilde x}_{{\tilde n}}) - {\tilde k}_{{\tilde n}-1}({\tilde x}_{{\tilde n}} - {\tilde x}_{{\tilde n}-1}) \nonumber \\
&\rightarrow& \frac{\partial}{\partial {\tilde n}}\left[ {\tilde k}_{{\tilde n}} \frac{\partial {\tilde x}_{{\tilde n}}}{\partial {\tilde n}}\right] \qquad ({\rm continuum \  limit})
\end{eqnarray}
and the viscous frictional force
\begin{eqnarray}
 {\tilde f}_{{\tilde n}}^{(vis)}= - {\tilde \gamma}_{{\tilde n}} \frac{\partial {\tilde x}_{{\tilde n}}}{\partial t}. 
\end{eqnarray}

Equation~(\ref{Eq.1}) indicates the expression for the local deformation 
\begin{eqnarray}
\frac{ \partial {\tilde x}_{{\tilde n}}}{\partial {\tilde n}} = \xi_{{\tilde n}}
\label{local_deformation}
\end{eqnarray}
One can then write Eq.~(\ref{f_balance}) as
\begin{eqnarray}
{\tilde k}_{{\tilde n}} \frac{\partial^2 {\tilde x}_{{\tilde n}}}{\partial {\tilde n}^2} + {\tilde \gamma}_{{\tilde n}} \frac{\partial {\tilde x}_{\tilde n}}{\partial t} = 0
\end{eqnarray} 
This line of argument was used to discuss the normal modes of the tethered chain stretched by flow~\cite{Macromolecules_Marciano_Brochard_1995}.

{\it Mapping to the p-Laplacian equation ---}

One may extrapolate the above force estimation at the blob scale to the segment scale in such a way that the elastic and viscous frictional forces acting on the $n$-th segment are given by $f_n^{(el)}={\tilde f}_{{\tilde n}}^{(el)}/g_{{\tilde n}} $ and $f_n^{(vis)}={\tilde f}_{{\tilde n}}^{(vis)}/g_{{\tilde n}}$.
The label index of blobs and that of segments are related as $n = \int_1^{{\tilde n}} g_{{\tilde n}'} d{\tilde n}'$.
We write respective forces as
\begin{eqnarray}
f_n^{(el)} = k_n \frac{\partial^2x_n}{\partial n^2} \\
f_n^{(vis)} = -\gamma_n \frac{ \partial x_n}{\partial t}
\end{eqnarray}
The spring and the friction constants $k_n$, $\gamma_n$ in this fine-grained frame can be estimated in the following way.
The relation $\partial n = g_{{\tilde n}} \partial {\tilde n}$ between internal coordinate before and after the fine-graining indicates the transformation rule of the local chain deformation
\begin{eqnarray}
\frac{\partial {\tilde x}_{{\tilde n}}}{\partial {\tilde n}} = g_{{\tilde n}}\frac{\partial x_n}{\partial n}
\label{transform_n}
\end{eqnarray}
This, together with Eq.~(\ref{local_deformation}), implies
\begin{eqnarray}
a \left( \frac{\xi_{\tilde n}}{a}\right)^{(\nu-1)/\nu} \simeq \frac{\partial x_n}{\partial n}.
\end{eqnarray}
These considerations lead to~\footnote{The second derivative relation also follows as $\partial^2 {\tilde x}_{{\tilde n}}/\partial {\tilde n}^2 = C g_{{\tilde n}}^2 \ \partial^2 x_n/\partial n^2$ with a negative coefficient $C=(\nu/(\nu-1)) <0$.}
\begin{eqnarray}
k_n \simeq k \left( \frac{\xi_{{\tilde n}}}{a}\right)^{(1-2\nu)/\nu} \simeq k\left(\frac{\partial x_n}{\partial (na)}\right)^{(2\nu -1)/(1-\nu)} \label{k_n_s}\\
\gamma_n \simeq \gamma \left( \frac{\xi_{{\tilde n}}}{a}\right)^{z-2-(1/\nu)} \simeq \gamma \left( \frac{\partial x_n}{\partial (na)}\right)^{[1-(z-2)\nu] /(1-\nu)}. \label{gamma_n_s}
\end{eqnarray}
The force balance relation $f_n^{(el)}+f_n^{(vis)}=0$ can be cast into a so-called {\it p}-Laplacian diffusion equation given in Eq.~(\ref{p_Lap})~\cite{EPL_Brochard_1994,EPL_Sebastian_2011}:

Again, useful insights can be deduced from the mono-block approximation~\cite{Macromolecules_Marciano_Brochard_1995}, where the blob sizes are assumed to be uniform with $\xi_{{\tilde n}} \simeq k_BT/f$
(see Sec.~\ref{scaling_theory}). 
In this approximation, the spring and the friction constants~Eqs.~(\ref{k_n_s}),\ (\ref{gamma_n_s}) do depend on $f$ but not on $n$:
\begin{eqnarray}
k_n = k^{(f)} \simeq k \left(\frac{fa}{k_BT} \right)^{(2\nu-1)/\nu} \label{k_f}\\
\gamma_n = \gamma^{(f)} \simeq \gamma \left(\frac{fa}{k_BT} \right)^{2-z+(1/\nu)} \label{gamma_f}
\end{eqnarray}
Therefore, Eq.~(\ref{p_Lap}) becomes a simple linear diffusion equation
\begin{eqnarray}
\frac{\partial x_n}{\partial t} = D_0 \left( \frac{fa}{k_BT}\right)^{z-(2/\nu)}\frac{\partial^2 x_n}{\partial (na)^2}
\label{p_Lap_f}
\end{eqnarray}
with the force dependent diffusion coefficient.

One can check that the self-similar scaling solution of Eq.~(\ref{p_Lap_f}) is consistent with the average drift of the tagged segment in strong force regime discussed in Sec.~\ref{scaling_theory}. Assume that at time $s$, the tension gets transmitted up to $m(s)$-th segments from the pulled end. 
Requiring Eq.~(\ref{p_Lap_f}) to be invariant under the scale transformation $t \rightarrow s t$ and $n \rightarrow n^*(s) n$,  one obtains the dynamics of the tension front $n^*(s)$, which is given by Eq.~(\ref{n*_f}).

Note that the above stretching process can also be analyzed by a different, but related nonlinear diffusion equation, which describes the time evolution of the segment line density field~\cite{PRE_Sakaue_2012, PRE_Rowghanian_Grosberg_2012, Macromolecules_Paturej_2012, PRE_Saito_2013}.

\section*{Appendix B}
\subsection*{Fluctuation-response relation in mode space}
We calculate the response function $\chi_p(t,s) \equiv \delta \langle {\dot X}_{p}(t) \rangle /\delta F_{p}(s)$ and the correlation function  $C_p(t,s)   \equiv \langle {\Delta \dot X}_{p}(t) {\Delta \dot X}_{p}(s)\rangle$ without assuming $t_0 \rightarrow -\infty$.
We first consider the case with unchanged spring and friction constants, which applies to the Rouse model, and the more general case with SA and HIs in the weak force regime. 

\subsubsection*{Weak force regime}

{\it Response function---}
Upon time derivative of Eq.~(\ref{Z_p_solution}) and taking ensemble average over the noise sequence $Z_p(t)$, the response function is obtained as
\begin{eqnarray}
\chi_p(t,s) = -\frac{k_p}{\gamma_p^2}e^{-(k_p /\gamma_p) (t-s)} + \frac{2}{\gamma_p}\delta(t-s).
\label{chi_p}
\end{eqnarray}

{\it Correlation function---}
The time correlation of $\Delta {\dot X}_p(t) \equiv {\dot X}_p(t) - \langle  {\dot X}_p(t) \rangle $ can be decomposed as 
\begin{eqnarray}
C_p(t,s; t_0) = C_{p}^{(st)}(t,s) + C_{p}^{(ex)}(t,s; t_0),
\label{C_p_total}
\end{eqnarray}
where the first is the stationary part invariant with respect to the time translation, i.e., $C_{p}^{(st)}(t,s)=C_{p}^{(st)}(t-s,0)$;
\begin{eqnarray}
C_{p}^{(st)}(t,s) =  -\frac{k_p}{\gamma_p^2} k_BT  e^{-(k_p /\gamma_p)(t-s)}  
+ \frac{2}{\gamma_p} k_BT \delta(t-s)
\label{C_p_st}
\end{eqnarray}
and the second is the excess due to the non-stationarity of the process;
\begin{eqnarray}
C_{p}^{(ex)}(t,s; t_0)  
= &-&\frac{k_p}{\gamma_p^2} k_BT e^{-(k_p /\gamma_p)(t+s-2t_0)} \nonumber \\
  &+&\frac{k_p^2}{\gamma_p^2} \langle X_p^2(t_0)\rangle e^{-(k_p /\gamma_p)(t+s-2t_0)},
\label{C_p_ex}
\end{eqnarray}
where we add an auxiliary argument $t_0$ to indicate the initial time.
One can verify the FDT~(\ref{FDT_p}), provided that the process is stationary, i.e., $t_0 \rightarrow -\infty$. Note that the excess part~(\ref{C_p_ex}) identically vanishes, when the equi-partition condition
\begin{eqnarray}
\langle X_p^2(t_0) \rangle = \frac{k_BT}{k_p}
\label{X_p_equilibrium}
\end{eqnarray}
holds for each of $p \ge 1$ modes at $t=t_0$, where the averaging is taken over the probability distribution of $X_p$ at $t=t_0$.

\subsubsection*{Strong force regime}

When the polymer with SA and/or HIs is stretched strongly, the calculation becomes slightly complicated due to the time dependence of spring and friction constants. 

{\it Response function---}
From the solution of Eq.~(\ref{Eq_NC_f}), the response function is obtained as
\begin{eqnarray}
&&\chi_{p,f}(t,s; \tau_{f0}) \nonumber \\
&&=  \left\{
\begin{array}{lll}
 \chi_p(t,s) & (s < t < \tau_{f0})  \\
 -\frac{k_p^{(f)}}{\gamma_p \gamma_p^{(f)}} e^{-(k_p/\gamma_p)(\tau_{f0}-s) -(k_p^{(f)}/\gamma_p^{(f)})(t-\tau_{f0})}  & (s<\tau_{f0} <t)  \\
  -\frac{k_p^{(f)}}{(\gamma_p^{(f)})^2}e^{-(k_p^{(f)} /\gamma_p^{(f)}) (t-s)} + \frac{2}{\gamma_p^{(f)}}\delta(t-s)  & (\tau_{f0}<s <t) .
\end{array}
\right.
\nonumber \\
\end{eqnarray}
For $s<t<\tau_{f0}$ case, the response function is the same as that for the weak force regime (Eq.~(\ref{chi_p})). For $\tau_{f0}<s<t$ case, it again takes the same functional form as Eq.~(\ref{chi_p}) with the replacement $(\gamma_p, k_p) \rightarrow (\gamma_p^{(f)}, k_p^{(f)})$. Only for the case $s<\tau_{f0}<t$, the stationarity in the response function breaks down, and there appears an auxiliary argument $\tau_{f0}$.

{\it Correlation function---}
From the solution of Eq.~(\ref{Eq_NC_f}), the fluctuation in the velocity $\Delta {\dot X}_{p}(t) \equiv {\dot X}_p(t) - \langle  {\dot X}_p(t) \rangle $ is obtained as
\begin{eqnarray}
\Delta {\dot X}_p(t) = 
   \left\{
\begin{array}{ll}
   \int_0^t ds \chi_p(t,s) Z(s) - X_p(0) \frac{\gamma_p}{k_p}e^{-(k_p/\gamma_p)t} & (\tau_{f0} > t) \\
  \int_{\tau_{f0}}^t ds \chi_{p,f}^{(\tau_{f0}<s<t)}(t,s;\tau_{f0}) Z_p^{(f)}(s) \nonumber \\
\qquad - \Delta X_p(\tau_{f0}) \frac{k_p^{(f)}}{\gamma_p^{(f)}}e^{-(k_p^{(f)}/\gamma_p^{(f)})(t-\tau_{f0})} & (t > \tau_{f0}) .
\end{array}
\right.
\\
\end{eqnarray}
From this, one can calculate the correlation of the velocity fluctuation $C_{p,f}(t,s; \tau_{f0}) = \langle \Delta {\dot X}_p(t) \Delta {\dot X}_p(s) \rangle$, and obtain the followings;
 (i) For $s<t<\tau_{f0}$, the weak force regime applies, so it is given by Eqs.~(\ref{C_p_total}) - (\ref{C_p_ex}); 
(ii) For $\tau_{f0}<s<t$, it can again be decomposed as
\begin{eqnarray}
C_{p,f}(t,s; \tau_{f0}) = C_{p,f}^{(st)}(t,s) + C_{p,f}^{(ex)}(t,s; \tau_{f0}),
\label{C_pf_total}
\end{eqnarray}
where $C_{p,f}^{(st)}(t,s)$ and $C_{p,f}^{(ex)}(t,s; \tau_{f0})$ take the same functional forms as those in the weak force regime (Eqs.~(\ref{C_p_st}) and~(\ref{C_p_ex}), respectively) with the replacement  $(\gamma_p, k_p,t_0) \rightarrow (\gamma_p^{(f)}, k_p^{(f)},\tau_{f0})$; (iii) For $s<\tau_{f0}<t$, it becomes
\begin{eqnarray}
&&C_{p,f}(t,s; \tau_{f0}) 
=   -\frac{k_p^{(f)} k_BT}{\gamma_p \gamma_p^{(f)}} e^{-(k_p/\gamma_p)(\tau_{f0}-s) -(k_p^{(f)}/\gamma_p^{(f)})(t-\tau_{f0})} \nonumber \\
 &-&  \frac{k_p^{(f)} }{\gamma_p \gamma_p^{(f)}} \left(k_BT - k_p \langle X_p(0)^2\rangle  \right)e^{-(k_p/\gamma_p)(\tau_{f0}+s) -(k_p^{(f)}/\gamma_p^{(f)})(t-\tau_{f0})}  
\nonumber  \\
\end{eqnarray}

\end{document}